\documentclass[aps,prd,preprintnumbers,nofootinbib,preprint]{revtex4}
\usepackage{bm}
\usepackage{latexsym}
\usepackage{dcolumn}
\usepackage{amsmath,amsfonts,amssymb}
\usepackage{graphicx,epsfig}
\usepackage{amsthm}
\usepackage{url,hyperref}
\usepackage{graphics}

\usepackage{amsmath,amssymb,dcolumn}
\usepackage{graphicx}
\usepackage{slashed}
\usepackage{subfigure}

\newcommand{\be}{\begin{eqnarray}}
\newcommand{\ee}{\end{eqnarray}}
\newcommand{\bea}{\begin{eqnarray}}
\newcommand{\eea}{\end{eqnarray}}

\def\comment#1{}

\usepackage{color}

\definecolor{darkred}{rgb}{.8,0,0}

\definecolor{darkblue}{rgb}{0,0,.7}

\definecolor{darkgreen}{rgb}{0,.7,0}

\begin{document}

%
%
\title{Joule-Thomson expansion for the regular(Bardeen)-AdS black hole}
%
%
%
%
%
\author{{Jin Pu}$^{1,2}$}\email[email:~]{pujin@cwnu.edu.cn}
\author{Sen Guo$^{2}$} \email[email:~]{quantumguosenedu@163.com}
\author{Qing-Quan Jiang$^{2}$}\email[email:~]{qqjiangphys@yeah.net}
\author{Xiao-Tao Zu$^{1}$\vspace{1ex}}\email[email:~]{xtzu@uestc.edu.cn}
\affiliation{$^1$School of Physics, University of Electronic Science and Technology of China, Chengdu 610054, China\vspace{1ex}}
\affiliation{$^2$College of Physics and Space Science, China West Normal University, Nanchong 637002, China\vspace{1ex}}

%
%
%
%
%
\begin{abstract}
%
%
%
%
%
%
\par\noindent
In this paper, we attempt to study the Joule-Thomson expansion for the regular black hole in an anti-de Sitter background, and obtain the inversion temperature and curve for the Bardeen-AdS black hole in the extended phase space. We investigate the isenthalpic and inversion curves for the Bardeen-AdS black hole in the $T-P$ plane to find the intersection points between them are exactly the inversion points discriminating the heating process from the cooling one. And, the inversion curve for the regular(Bardeen)-AdS black hole is not closed and there is only a lower inversion curve in contrast with that of the Van der Walls fluid. Most importantly, we find the ratio between the minimum inversion and critical temperature for the regular(Bardeen)-AdS black hole is 0.536622, which is always larger than all the already-known ratios for the singular black hole. This larger ratio for the Bardeen-AdS black hole in contrast with the singular black hole may stem from the fact that there is a repulsive de Sitter core near the origin of the regular black hole.\\
\end{abstract}
%
%
%
%
\maketitle
%
%
%
%
%
%
\section{Introduction}
\label{intro}
%
%
\par\noindent
It is well known that a black hole as a thermodynamic system has many interesting consequences. It sets deep and fundamental connections between the laws of classical thermodynamics, quantum mechanics and general relativity. The thermodynamic properties of black holes have been extensively investigated since the first studies of Bekenstein and Hawking \cite{1,2,3,4,5,6}. Black holes as thermodynamic systems have many similarities with general thermodynamic systems. For the black holes in the AdS space, these similarities become more obvious and precise. Hawking and Page have pioneered the study of the phase transition between the Schwarzschild-AdS black hole and the thermal AdS space \cite{7}. Then, Chamblin and Emparan have observed the similar relation between the charged AdS black holes and the Van der Waals liquid-gas systems in \cite{8,9}.
\par
Recently, black hole thermodynamics in the AdS space have been intensively studied in the extended phase space where the cosmological constant is considered as the thermodynamic pressure
\be\label{1-1}
P=-\frac{\Lambda}{8\pi},
\ee
and its conjugate quantity as the thermodynamic volume\cite{10,11}
\be\label{1-2}
V={\Big(\frac{\partial M}{\partial P}\Big)}_{S,Q,J},
\ee
which can enrich the AdS black hole thermodynamics. And a lot of research have shown that the charged AdS black holes is similar to the Van der Waals liquid-gas in the extended phase space \cite{12}, and the $P-V$ criticality of the black hole systems behaves like that of Van der Waals systems \cite{12,13,14,15,16,17,18,19,20,21,22,23,24,25,26,27,28,29,30,31}. In \cite{32,33,34,35}, it can also be found that black holes have the second-order and third-order phase transitions in addition to the Van der Waals-like phase transition.
\par
Apart from the phase transitions and critical phenomena, the thermodynamic analogies between the AdS black holes and the Van der Waals systems have been creatively generalized to the well-known Joule-Thomson expansion process in recent months. In classical thermodynamics, the Joule-Thomson expansion is gas at a high pressure passing through a porous plug to a section with a low pressure in which enthalpy is a constant. In \cite{36}, \"{O}kc\"{u} and Aydmer have first investigated the Joule-Thomson expansion of the charged RN-AdS black hole in the extended phase space, where the inversion and isenthalpic curves are obtained and the heating-cooling regions are shown in the $T-P$ plane. Subsequently, these pioneering works have been generalized to some kinds of black holes, such as quintessence charged AdS black hole \cite{37}, Kerr-AdS black hole \cite{38}, d-dimensional charged AdS black holes \cite{39}, holographic super-fluids \cite{40}, charged AdS black hole in $f(r)$ gravity \cite{41}, AdS black hole with a global monopole \cite{42}, AdS black holes in Lovelock gravity \cite{43}, charged Gauss-Bonnet black holes \cite{44}, Ads black hole in Einstien-Maxwell-axions theory and AdS black hole in massive gravity\cite{45}.
\par
However, the above-mentioned Joule-Thomson expansion processes are only for the AdS black holes with singularity. For the regular black hole without singularity, the Joule-Thomson expansion has not been discussed so far. Bardeen has obtained the first solution of the regular black holes with nonsingular geometry satisfying the weak energy condition \cite{46}. Subsequently, Ay\'{o}n-Beato and Garc\'{\i}a have proved that the Bardeen black hole can be interpreted as a gravitationally collapsed magnetic monopole arising in a specific form of non-linear electrodynamics \cite{47}. Then, the stabilities of the Bardeen-like regular black holes have been studied by Moreno and Sarbach\cite{48}, and the thermodynamic properties of these regular black holes have been also discussed in \cite{49,50,51,52,53,54,55,56,57}.  Recently, Tzikas has found the Bardeen-AdS black hole thermodynamic system appears criticality and a first order small/large black hole phase transition, and also found the critical exponents is exactly the same as those of the Van der Waals gas \cite{58}. Obviously, the phase transition and the $P-V$ criticality of the Ads black hole are independent of the spacetime with or without singularity. In this paper, taking the regular(Bardeen)-AdS black hole as an example, we attempt to observe the dependence of the spacetime without singularity on the Joule-Thomson expansion.
\par
The remainders of this paper are organized as follows. In Sec.2, we review the thermodynamic properties of the Bardeen-AdS black hole in the extended phase space, and obtain the equation of state for the black hole. In Sec.3, we apply the Joule-Thomson expansion for the Van der Waals fluid and the Bardeen-AdS black hole, and plot the inversion and isenthalpic curves to determine the cooling and heating regions in the $T-P$ plane. Sec.4 ends up with some conclusions. Here we use the units $G_{N}=\hbar=\kappa_{B}=c=1$.
%
%
\section{The thermodynamics of the Bardeen-AdS black hole}
\label{sec2}
%
%
%
\par\noindent
In this section, we briefly review the thermodynamic properties of the Bardeen-AdS black hole in the extended phase space. According to the extension proposed in \cite{47}, the corresponding action with a negative $\Lambda$-term is expressed as \cite{58}
\be\label{2-1}
\mathcal{S}=\frac{1}{16\pi}\int d^{4}x \sqrt{-g}\Big(R+\frac{6}{l^{2}}-4{\pounds}(F)\Big),
\ee
where $R$ and $g$ are the Ricci scalar and the determinant of the metric tensor, respectively. And $l$ is the positive AdS radius, which is related to $\Lambda$ by the relation $\Lambda=-3/l^{2}$. $\pounds(F)$ is a function of $F=\frac{1}{4}F^{\mu\nu}F_{\mu\nu}$, which can be given by
\be\label{2-2}
\pounds(F)=\frac{2M}{q^{3}}{\Big(\frac{\sqrt{4F q^{2}}}{1+\sqrt{4F q^{2}}}\Big)}^{\frac{5}{2}},
\ee
where $q$ and $F_{\mu\nu}$ are the charge and the field tensor, respectively. The line element of the Bardeen-AdS black hole is given by \cite{58,59}
\be\label{2-3}
ds^{2}=-{\Big(1-\frac{2m(r)}{r}\Big)}dt^{2}+{\Big(1-\frac{2m(r)}{r}\Big)}^{-1}dr^{2}+r^{2}{d\Omega}^{2},
\ee
where
\be\label{2-4}
m(r)=\frac{Mr^{3}}{{(q^{2}+r^{2})}^{\frac{3}{2}}}-\frac{r^{3}}{2l^{2}},
\ee
and ${d\Omega}^{2}={d\theta}^{2}+{\sin}^{2}{\theta}{d\phi}^{2}$. The Bardeen-AdS metric potential can be expressed as
\be\label{2-5}
f(r)=1-\frac{2 M r^{2}}{{(q^{2}+r^{2})}^{\frac{3}{2}}}+\frac{r^{2}}{l^{2}},
\ee
where $M$ is the mass of the black hole. One can obtain the event horizon $r_h$ of the black hole by the largest root of $f(r_{h})=0$. From Eq.({\ref{2-5}}), the mass $M$ of the black hole can be written by
\be\label{2-6}
M=\frac{(l^{2}+{r^2_{h}}){(q^{2}+{r^2_{h}})}^{\frac{3}{2}}}{2l^{2}{r^2_{h}}}.
\ee
\par
In the extended phase space, the cosmological term allows for the definition of the pressure (\ref{1-1}) of the system along with its conjugate quantity\cite{10,13}, so the black hole pressure is expressed as
\be\label{2-7}
P=\frac{3}{8\pi l^{2}}.
\ee
The entropy of the black hole at the horizon is given by
\be\label{2-8}
S=\pi r^2_h.
\ee
We consider the mass $M$ of the black hole as the functions of entropy $S$, pressure $P$ and charge $q$, i.e. $M=M(S,P,q)$ (see Eqs.(\ref{2-6}), (\ref{2-7}) and (\ref{2-8})), and then the mass $M$ of the black hole can be reformulated as
\be\label{2-9}
M=\frac{(\pi q^2+S)^{3/2}(3+8PS)}{6\sqrt{\pi}S}.
\ee
\par
The first law of the black hole is given by
\be\label{2-10}
dM=TdS+VdP+{\varphi}{dq},
\ee
and all the thermodynamic variable defined above can be obtained as follows
\be\label{2-11}
T=\Big(\frac{\partial M}{\partial S}\Big)_{P,q},~~V=\Big(\frac{\partial M}{\partial P}\Big)_{S,q},~~\varphi={(\frac{\partial M}{\partial q})}_{S,P}.
\ee
According to Eqs.(\ref{2-9}) and (\ref{2-11}), the temperature of the black hole is obtained as
\be\label{2-12}
T=\frac{\sqrt{q^2+S/\pi}(S+8PS^2-2\pi q^2)}{4S^2},
\ee
and the black hole volume is given by
\be\label{2-13}
V=\frac{4\pi}{3}(q^2+\frac{S}{\pi})^{\frac{3}{2}}.
\ee
Similarly, we have
\be\label{2-14}
{\varphi}=\frac{\pi q(3+8PS)\sqrt{q^2+S/\pi}}{2S}.
\ee
One can combine these thermodynamical quantities presented above to obtain the Smarr relation, which can be expressed as
\be\label{2-15}
M=2TS-2PV+\varphi q.
\ee
\par
Moreover, combining the expressions for the thermodynamic pressure (\ref{2-7}), the entropy (\ref{2-8}) and the temperature (\ref{2-12}), we get the equation of state for the black hole as follows
\be\label{2-16}
P=\frac{T}{2\sqrt{q^2+r^2_{h}}}-\frac{1}{8 \pi r^2_{h}}+\frac{q^{2}}{4\pi {r^4_{h}}}.
\ee
At the critical point, we have \cite{62}
\be\label{2-17}
\frac{\partial P}{\partial r_{h}}=\frac{{\partial}^{2} P}{{\partial}{r^2_{h}}}=0,
\ee
so the critical temperature $T_c$ is obtained as
\be\label{2-18}
T_{c}=\frac{25\sqrt{31+13\sqrt{10}}}{432(5+2\sqrt{10})^{\frac{3}{2}}\pi q}.
\ee
\par
In this section, we have presented some thermodynamic quantities of the Bardeen-AdS black hole in the extended space. In the next section, we will use these quantities to investigate the Joule-Thomson expansion for the black hole.
%
%
\section{The Joule-Thomson expansion}
\label{sec3}
%
%
%
\par\noindent
In this section, we study the well-known Joule-Thomson expansion of the Bardeen-AdS black hole compared with the expansion of the Van der Waals fluid. The Joule-Thomson expansion is a classic physical process with two important characteristics that the temperature changes with pressure and enthalpy are constant during the expansion process \cite{62}. So the Joule-Thomson expansion of a black hole is an isenthalpy process in the extended phase space, and the Joule-Thomson coefficient $\mu$, whose sign can be utilized to determine whether heating or cooling will occur, is defined via the change of the temperature with respect to the pressure as follows
\begin{eqnarray}\label{3-1}
\mu={\Big(\frac{\partial T}{\partial P}\Big)}_{H}.
\end{eqnarray}
\par
Since the pressure always decreases during the expansion process, the change of the pressure is negative, which may cause the temperature to decrease or increase during the process. Therefore, the change of the temperature determines the sign of $\mu$. If $\mu$ is negative (positive), heating (cooling) occurs, so the gas warms (cools). In the extended phase space, we compare a black hole system to a Van der Waals fluid system with a fixed number of particles, we should here consider canonical ensemble with fixed change $q$. The Joule-Thomson coefficient is then given by \cite{36}
\be\label{3-2}
\mu={\Big(\frac{\partial T}{\partial P}\Big)}_{M}=\frac{1}{C_p}\Big[T\Big({\frac{\partial V}{\partial T}}\Big)_P-V\Big].
\ee
Setting $\mu=0$, we can obtain the inversion temperature, i.e.
\be\label{3-3}
T_{i}=V {\Big(\frac{\partial T}{\partial V}\Big)}_{P}.
\ee
\par
In the next subsection, we use the Joule-Thomson coefficient (\ref{3-1}) and (\ref{3-2}) to determine the inverse curves for the Van der Waals fluid and the Bardeen-AdS black hole in the extended phase space. At the same time, the inversion temperatures for them are obtained from Eq.(\ref{3-3}).
%
%
\subsection{The Van der Waals fluid}
\label{sec3-1}
%
%
\par\noindent
It is well-known that the Van der Waals equation is an improvement of the ideal gas equation, which is characterized by taking into account the size of gas molecules and the interactions between molecules that are ignored by the ideal gas, so as to better describe the gas-liquid phase transition behaviors of the actual fluids \cite{63,64}. The Van der Waals equation is expressed as
\be\label{3-1-1}
P=\frac{k_{B}T}{\nu-b}-\frac{a}{\nu^{2}},
\ee
where $\nu=V/N$, $P$, $T$ and $k_{B}$ denote the specific volume, pressure, temperature, and Boltzmann constant, respectively. Constants $a$ and $b$ are respectively measures of the attraction between the molecules and the average space occupied by each molecule, which can be determined from experimental data.
\par
\begin{figure}
\centering\includegraphics[width=0.4\textwidth]{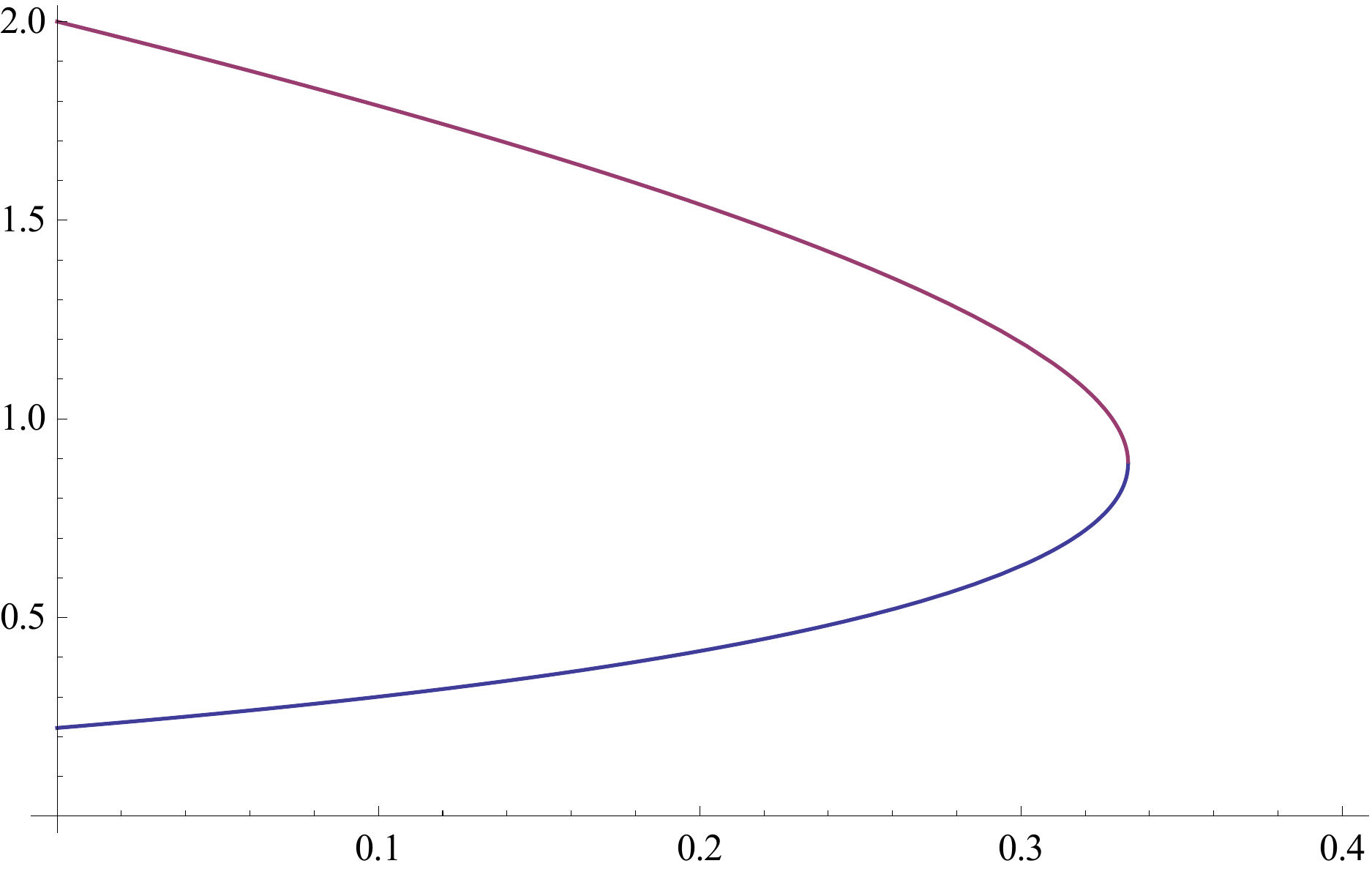}
\caption{The inversion curves for the Van der Waals fluid. The red(blue) line corresponds to the lower(upper) inversion curve.}
\label{fig1}
\end{figure}

In order to further study the Joule-Thomson expansion, it is useful to give some thermodynamic properties of the Van der Waals equation. The internal energy of the Van der Waals fluid is given by \cite{65}
\be\label{3-1-2}
U=\frac{2}{3}k_{B} T-\frac{a}{\nu},
\ee
and then we can get the enthalpy as
\be\label{3-1-3}
H(T,\nu)=\frac{2}{3}k_{B}T+\frac{k_{B}T \nu}{\nu-b}-\frac{2a}{\nu}.
\ee
Using to Eq.(\ref{3-3}), the inversion temperature is obtained as
\be\label{3-1-4}
T_{i}=\frac{1}{k_{B}}\Big(P_{i}\nu-\frac{a}{\nu}+\frac{2ab}{\nu^{2}}\Big),
\ee
where $P_i$ denotes the inversion pressure. According to the Van der Waals equation ({\ref{3-1-1}}), one can have
\be\label{3-1-5}
T_{i}=\frac{1}{k_{B}}\Big(P_{i}\nu-P_{i}b+\frac{a}{\nu}-\frac{ab}{\nu^{2}}\Big),
\ee
Subtracting Eq.({\ref{3-1-4}}) from Eq.({\ref{3-1-5}}), it yields
\be\label{3-1-6}
Pb\nu^{2}-2a\nu+3ab=0.
\ee
Solving the above equation (\ref{3-1-6}) for $\nu(P_i)$, we can get two roots as follows
\be
\nu=\frac{a\pm\sqrt{a^2-3ab^2P_i}}{bP_i}.
\ee
If we substitute the two roots into Eq.(\ref{3-1-5}), we can obtain the inversion temperature as
\be\label{3-1-7}
T_{i}=\frac{2(5a-3b^{2}P_{i}\pm 4\sqrt{a^{2}-3ab^{2}P_{i}})}{9bk_{B}}.
\ee
The positive and negative signs in the above equation (\ref{3-1-7}) give upper and lower inversion curves, respectively. In Fig.\ref{fig1}, we plot the inversion curves of the Van der Waals system in the $T-P$ plane basing on Eq. (\ref{3-1-7}) and determine the cooling-heating regions.
\par
Setting $P_{i}=0$, the minimum and maximum inversion temperatures can be obtained as
\be\label{3-1-8}
T^{min}_i=\frac{2a}{9bk_{B}},~~~T^{max}_i=\frac{2a}{bk_{B}}.
\ee
The critical temperature of the Van der Waals fluid is given by $T_{c}=8a/(27bk_B)$, the ratio between the inversion and critical temperature is given by
\be\label{3-1-9}
\frac{T^{min}_i}{T_c}=\frac{3}{4},~~~\frac{T^{max}_i}{T_c}=\frac{27}{4}.
\ee
%
%
\subsection{The Bardeen-AdS black hole}
\label{sec3-2}
%
%
\par\noindent
In this subsection, we investigate the Joule-Thomson expansion of the Bardeen-AdS black hole in the extended phase space. The characteristic of the expansion is that temperature changes with pressure, and enthalpy remains constant during the expansion process. We know from \cite{13} that the mass of the black hole is determined as enthalpy in the AdS space, so the mass of the black hole remains constant during the expansion process. And in our case, by observing carefully the expressions of these thermodynamic quantities, we find a simpler method to derive the Joule-Thomson coefficient $\mu$. From Eqs.(\ref{2-7})(\ref{2-8}) and (\ref{2-9}), the pressure $P$ can be rewritten as a function of ${M}$ and $r_h$, i.e.
\be\label{3-2-1}
P({M},{r}_{h})=\frac{6M{r^2_{h}}-3{(q^{2}+{r^2_{h}})}^{\frac{3}{2}}}{8\pi{r^2_{h}}(q^{2}+r^2_h)^{\frac{3}{2}}},
\ee
then substituting $P(M,r_h)$ into the temperature (\ref{2-12}), it yields
\be\label{3-2-2}
T({M},{r}_{h})=\frac{3Mr^4_h-(q^2+r^2_h)^{\frac{5}{2}}}{2\pi r^4_h(q^2+r^2_h)^{\frac{5}{2}}}.
\ee
From these relations between the Joule-Thomson coefficient $\mu$, the pressure $P({M},{r}_{h})$ and the temperature $T({M},{r}_{h})$, one can obtain a simple expression as
\be\label{3-2-3}
\mu={\Big(\frac{\partial T}{\partial P}\Big)}_{M}={\Big(\frac{\partial T}{\partial {r_{h}}}\Big)}_{M}{\Big(\frac{\partial {r_{h}}}{\partial P}\Big)}_{M}=\frac{\Big(\partial T/\partial {r_{h}\Big)}_{M}}{\Big(\partial P/\partial {r_{h}}\Big)_{M}}.
\ee
Then, setting $\mu=0$ in Eq.(\ref{3-2-3}), one can get
\be\label{3-2-4}
8\pi Pr^6_h+2r^4_h-5q^2r^2_h-4q^4=0.
\ee
By solving the equation (\ref{3-2-4}) for $r_h(P_i)$, one can get a positive and real root which has a physically meaning, and the corresponding pressure $P_i$ in the expression of the root is the inversion pressure. Then substituting the root into the temperature expression (\ref{2-12}), one can obtain the inversion temperature $T_i$. Since these expressions are too complex, it is not specifically written here.
\par
\begin{figure}
\centering\includegraphics[width=0.4\textwidth]{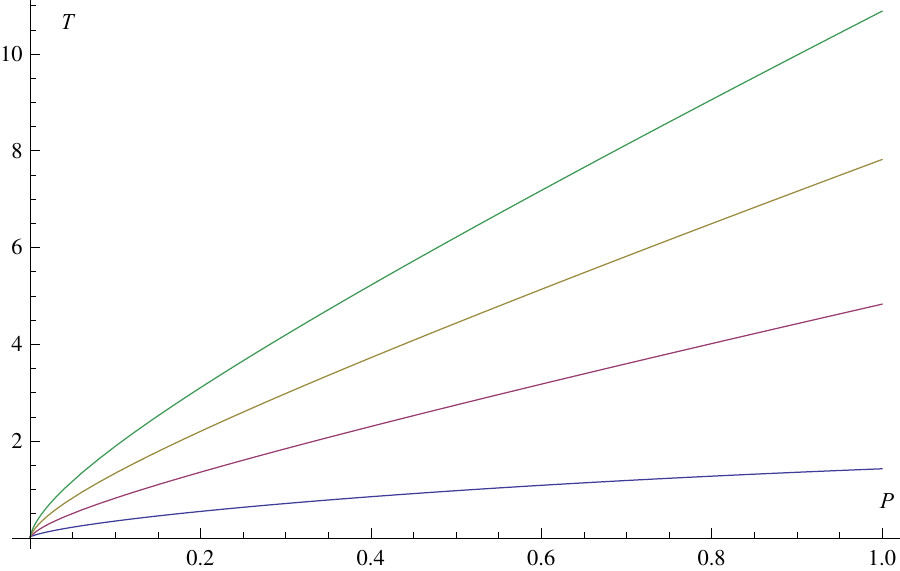}
\caption{The inversion curves of the Bardeen-AdS black hole in the $T-P$ plane. From bottom to top, the inversion curves correspond to $q=1,2,3,4$.}
\label{fig2}
\end{figure}

At the same time, we can also draw the inversion temperature curve according to the expression of the inversion temperature $T_i$, as shown in Fig.~\ref{fig2}. In Fig.~\ref{fig2}, the inversion curves of the Bardeen-AdS black hole are presented for various values of charge $q$, and the inversion temperature for a given pressure tends to increase with an increasing of $q$. Comparing with the inversion curve of the Van der Waals fluid, we can also find the inversion curves of the black hole are not closed and there is only a lower inversion curve, it means that the black holes always cool above the inversion curve during the Joule-Thomson expansion process, as previously described in \cite{36,37,38,39,40,41,42,43,44,45}.

\par
If we set the inversion pressure $P_i$ to be equal to zero, the minimum inversion temperature is given by
\be\label{3-2-6}
{T_{i}}^{min}=\frac{\sqrt{9+\sqrt{57}}(\sqrt{57}-3)}{2(5+\sqrt{57})^2 \pi q}.
\ee
The ratio between the minimum inversion and critical temperature is given by
\be
\frac{{T_{i}}^{min}}{T_{c}}{\approx} 0.536622.
\label{3-2-7}
\ee
This ratio between the minimum inversion and critical temperature for the Bardeen-AdS black hole is smaller than 0.75 for the Van der Waals fluid, but is larger than all the already-known ratios for the singular black holes. This larger ratio is universal for the regular black hole in contrast to the singular black hole. Therefore, there should be some obvious properties for the regular black hole. In fact, there is a repulsive de Sitter core near the origin of the regular black hole, however there is not for the singular black hole (see in Sec.4).

\begin{figure}
\centering
\subfigure[q=1 and M=2,2.5,3,3.5]
{\includegraphics[width=0.45\textwidth]{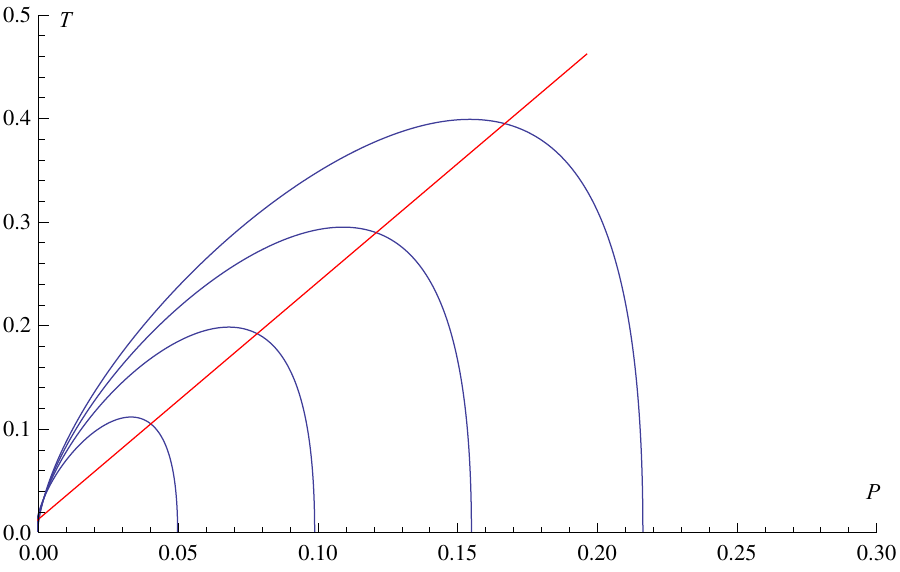}}
\subfigure[q=2 and M=5,5.5,6,6.5]
{\includegraphics[width=0.45\textwidth]{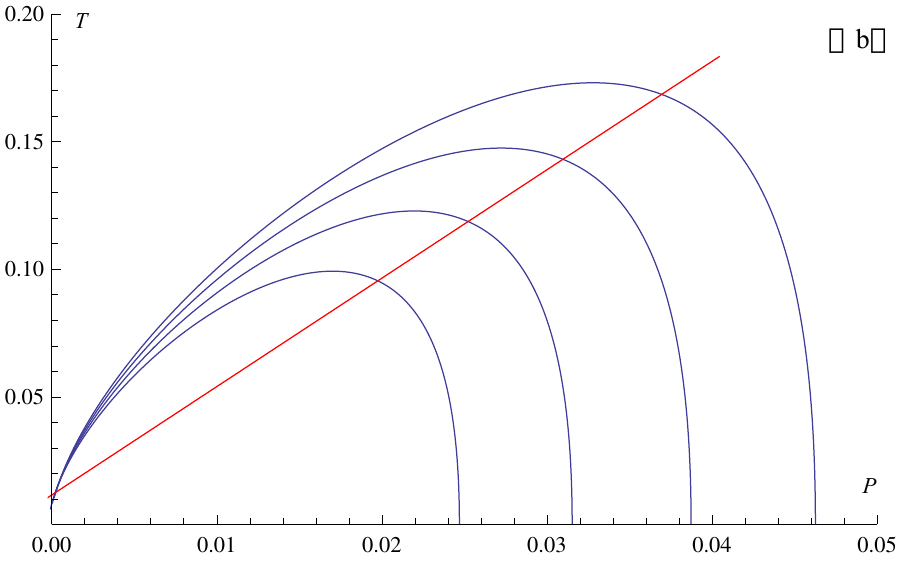}}
\subfigure[q=3 and M=8,8.5,9,9.5]
{\includegraphics[width=0.45\textwidth]{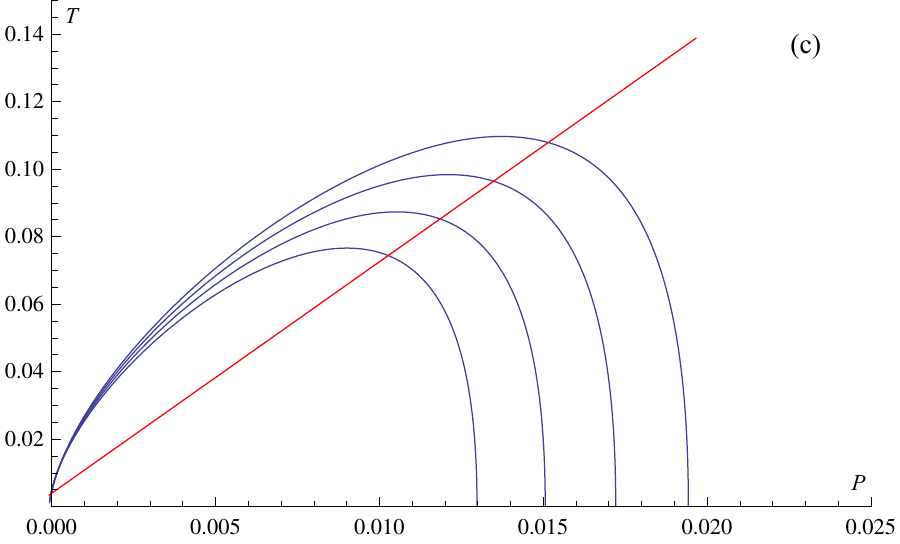}}
\subfigure[q=4 and M=12,12.5,13,13.5]
{\includegraphics[width=0.45\textwidth]{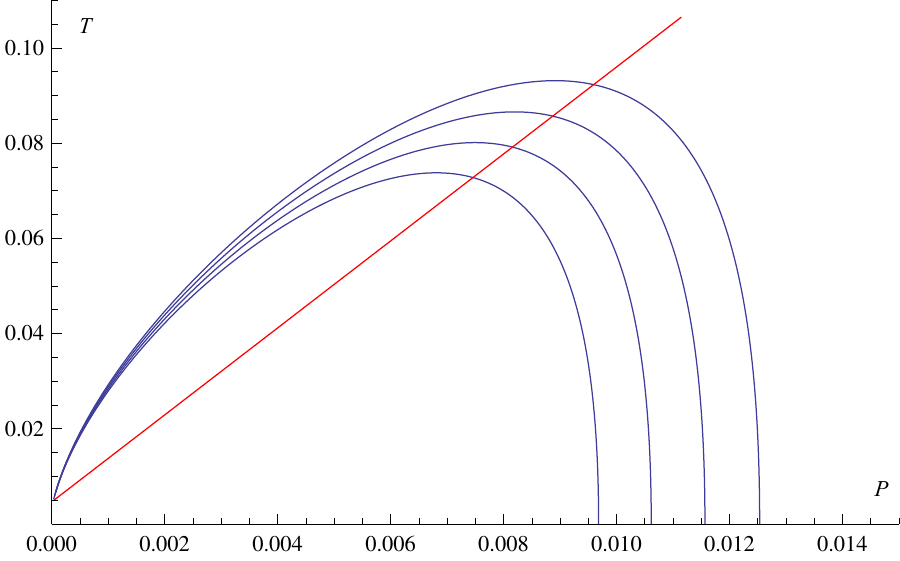}}
\caption{The inversion and isenthalpic(constant mass) curves of the Bardeen-AdS black hole. The blue (black) lines present the inversion (isenthalpic) curves. From bottom to top, the isenthalpic curves correspond to the increasing values of the mass ${M}$ of the black hole.}
\label{fig3}
\end{figure}
\par
In Fig.\ref{fig3}, we plot the isenthalpic (constant mass) and inversion curves for various values of charge $q$ in the $T-P$ plane. We can see the inversion curves dividing the plane into two regions. The region above the inverse curves corresponds to the cooling region, while the region under the inversion curves corresponds to the heating region. In fact, the heating and cooling regions are checked by the slope signs of the isenthalpic curve. The sign of slope is positive in the cooling region and changes in the heating region. In addition, the inversion curve acts as a boundary between the two regions, and cooling (heating) does not occur on the inversion curve.
%
%
\section{Conclusions and Discussions}
\label{sec4}
%
%
%
\par\noindent
In this paper, we have investigated the Joule-Thomson expansion for the Bardeen-AdS black hole in the extended phase space, where the cosmological constant is viewed as pressure and the black hole mass as enthalpy. Firstly, we have reviewed the thermodynamic properties of the Bardeen-AdS black hole in the extended phase space, and obtained the equation of state for the black hole. Then, by applying the Joule-Thomson expansion for the Van der Waals fluid, we have plotted the inversion curves and determined the cooling and heating regions in the $T-P$ plane. At the same time, we have investigated the Joule-Thomson expansion of the Bardeen-AdS black hole, and also plotted the inversion curves. In contrast to the Van der Waals fluid, the inversion curves of the Bardeen-AdS black hole are not closed and there is only a lower inversion curve. In addition, we have plotted the inversion and isenthalpic curves of the Bardeen-AdS black hole in the $T-P$ plane, and determined the cooling and heating regions for various values of charge $q$ and mass $M$. It shown the isenthalpic curves have positive slopes above the inversion curves so cooling occurs here, and the signs of slopes change under the inversion curves and heating occurs in this region.
\par
This ratio between the minimum inversion and critical temperature for the Bardeen-AdS black hole is smaller than 0.75 for the Van der Waals fluid, but is larger than all the already-known ratios for the singular black holes. This larger ratio is universal for the regular black hole in contrast to the singular black hole. Therefore, there should be some obvious properties for the regular black hole. In fact, at the large distances ($r\gg q$), the Bardeen-AdS black hole coincides with the conventional Schwarzschild-AdS black hole with $f(r)\approx 1-\frac{2M}{r}+\frac{r^2}{l^2}$. Near the origin ($r\ll q$), the Bardeen-AdS black hole could be described by the line element where $f(r)\approx 1-\Big(\frac{2Ml^2}{q^3}-1\Big)\frac{r^2}{l^2}$. According to this metric potential, the origin can be seen as: \textbf{\textit{i)}} a repulsive de Sitter core when $2Ml^2>q^3$; \textbf{\textit{ii)}} an attractive Anti de Sitter core when $2Ml^2<q^3$; \textbf{\textit{iii)}} a local Minkowski core with no gravitational interaction when $2Ml^2=q^3$. Obviously, when the origin is an attractive Anti de Sitter core or a local Minkowski core, there is no horizons. Therefore, the Bardeen-AdS black hole exists only in the case of a repulsive de Sitter core at the origin. Anyway, there is a repulsive de Sitter core for the regular(Bardeen)-AdS black hole in contrast with the singular black hole. Thus, we conclude the larger ratio for the regular(Bardeen)-AdS black hole in contrast with the singular black hole may stem from the fact that there is a repulsive de Sitter core near the origin of the regular black hole. On the other hand, for the Van der Waals fluid, there is also a repulsive volume in contrast with the ideal fluid, and much work has shown the ratio between the minimum inversion and critical temperature is always larger than that of the singular black hole. Therefore, it is reasonable to consider that a repulsive de Sitter core near the origin of the regular(Bardeen)-AdS black hole leads to the larger ratio in contrast with the singular black hole.
%
%
\section{Acknowledgements}
\par\noindent
This work is supported by the Program for NCET-12-1060, by the Sichuan Youth Science and Technology Foundation with Grant No. 2011JQ0019, and by FANEDD with Grant No. 201319, and by Ten Thousand Talent Program of Sichuan Province, and by Sichuan Natural Science Foundation with Grant No. 16ZB0178, and by the starting funds of China West Normal University with Grant No.17YC513 and No.17C050. After we have completed this paper, a similar reference was appeared in arXiv:1904.09548.
%
%

%
%
\end{document}